\documentclass[twocolumn,aps,amsmath,amssymb,floatfix,showpacs,superscriptaddress]{revtex4}
\usepackage{amsmath,amssymb,eucal,graphicx,bm}
\usepackage{graphicx}
\begin{document}
\title{Randomness in Competitions}
\author{E.~Ben-Naim}
\affiliation{Theoretical Division and Center for Nonlinear Studies,
Los Alamos National Laboratory, Los Alamos, New Mexico 87545 USA}
\author{N.~W.~Hengartner}
\affiliation{Computing and Computer Science Division, Los Alamos
National Laboratory, Los Alamos, New Mexico 87545 USA}
\author{S.~Redner}
\affiliation{Department of Physics, Boston University, Boston,
Massachusetts, 02215 USA}
\author{F.~Vazquez}
\affiliation{Max-Planck-Institut f\"ur Physik Komplexer Systeme,
N\"othnitzer Str.~38, 01187 Dresden, Germany}
\begin{abstract}
  We study the effects of randomness on competitions based on an
  elementary random process in which there is a finite probability
  that a weaker team upsets a stronger team.  We apply this model to
  sports leagues and sports tournaments, and compare the theoretical
  results with empirical data.  Our model shows that
  single-elimination tournaments are efficient but unfair: the number
  of games is proportional to the number of teams $N$, but the
  probability that the weakest team wins decays only algebraically
  with $N$.  In contrast, leagues, where every team plays every other
  team, are fair but inefficient: the top $\sqrt{N}$ of teams remain
  in contention for the championship, while the probability that the
  weakest team becomes champion is exponentially small. We also
  propose a gradual elimination schedule that consists of a
  preliminary round and a championship round. Initially, teams play a
  small number of preliminary games, and subsequently, a few teams
  qualify for the championship round.  This algorithm is fair and
  efficient: the best team wins with a high probability and the number
  of games scales as $N^{9/5}$, whereas traditional leagues require
  $N^3$ games to fairly determine a champion.
\end{abstract}
\maketitle

\section{Introduction}

Competitions play an important role in society \cite{csv,btd,br,msk},
economics \cite{ra}, and politics.  Furthermore, competitions underlie
biological evolution and are replete in ecology, where species compete for
food and resources \cite{sjg}.  Sports are an ideal laboratory for studying
competitions \cite{pn,pjs,hmr,fr}.  In contrast with evolution, where records
are incomplete, the results of sports events are accurate, complete, and
widely available \cite{abc,gts}.

Randomness is inherent to competitions. The outcome of a single match
is subject to a multitude of factors including game location, weather,
injuries, etc, in addition to the inherent difference in the strengths
of the opponents.  Just as the outcome of a single game is not
predictable, the outcome of a long series of games is also not
completely certain. In this paper, we review a series of our studies
that focus on the role of randomness in competitions
\cite{bvr,bvr1,brv,bh}. Among the questions we ask are: What is the
likelihood that the strongest team wins a championship?  What is the
likelihood that the weakest team wins?  How efficient are the common
competition formats and how ``accurate'' is their outcome?

We introduce an elementary model where a weaker team wins against a stronger
team with a fixed \emph{upset probability} $q$, and use this elementary
random process to analyze a series of competitions \cite{bvr}.  To help
calibrate our model, we first determine the favorite and the underdog from
the win-loss record over many years of sports competition from several major
sports.  We find that the distribution of win percentage approaches a
universal scaling function when the number of games and the number of teams
are both large. We then simulate a realistic number of games and a realistic
number of teams, and demonstrate that our basic competition process
successfully captures the empirical distribution of win percentage in
professional baseball \cite{bvr1}.  Moreover, we study the empirical upset
frequency and observe that this quantity differentiates professional sports
leagues, and furthermore, illuminates the evolution of competitive balance.

Next, we apply the competition model to single-elimination tournaments where,
in each match, the winner advances to the next round and the loser is
eliminated \cite{brv}.  We use the very same competition rules where the
underdog wins with a fixed probability.  Here, we introduce the notion of
innate strength and assume that entering the competition, the teams are
ranked.  We find that the typical rank of the winner decays algebraically
with the size of the tournament.  Moreover, the rank distribution for the
winner has a power-law tail.  Hence, larger tournaments do produce stronger
winners, but nevertheless, even the weakest team may have a realistic chance
of winning the entire tournament.  Therefore, tournaments are efficient but
unfair.

Further, we study the league format, where every team plays every other team
\cite{bh}.  We note that the number of wins for each team performs a biased
random walk. Using heuristic scaling arguments, we establish that the top
$\sqrt{N}$ teams have a realistic chance of becoming champion, while it is
highly unlikely that the weakest teams can win the championship. In addition,
the total number of games required to guarantee that the best team wins is
cubic in $N$. In this sense, leagues are fair but inefficient.

Finally, we propose a gradual elimination algorithm as an efficient way to
determine the champion. This hybrid algorithm utilizes a preliminary round
where the teams play a small number of games and a small fraction of the
teams advance to the next round.  The number of games in the preliminary
round is large enough to ensure the stronger teams advance. In the
championship round, each team plays every other team ample times to guarantee
that the strongest team always wins. This algorithm yields a significant
improvement in efficiency compared to a standard league schedule.

The rest of this paper is organized as follows. In section II, the basic
competition model is introduced and its predictions are compared with
empirical standings data. The notion of innate team strength is incorporated
in section III, where the random competition process is used to model
single-elimination tournaments. Scaling laws for the league format are
derived in section IV.  Scaling concepts are further used to analyze the
gradual elimination algorithm proposed in section V. Finally, basic features
of our results are summarized in section VI.

\section{The competition model} 

In our competition model, $N$ teams participate in a series of games. Two
teams compete head to head and, at the end of each match, one team is
declared the winner and the other as the loser. There are no ties.

To study the effect of randomness on competitions, we consider the scenario
where there is a fixed \emph{upset probability} $q$ that a weaker team upsets
a stronger team \cite{btd,bvr}.  This probability has the bounds $0\leq q
\leq 1/2$.  The lower bound corresponds to predictable games where the
stronger team always wins, and the upper bound corresponds to random
games. We consider the simplest case where the upset probability $q$ does not
change with time and is furthermore independent of the relative strengths of
the competitors.

In each game, we determine the stronger and the weaker team from 
current win-loss records.  Let us consider a game between a team with
$k$ wins and a team with $j$ wins. The competition outcome is 
stochastic: if $k>j$,
\begin{equation}
\label{compete}
(k,j)\to
\begin{cases}
(k+1,j)\quad{\rm with\ probability\ } p,\\  
(k,j+1)\quad{\rm with\ probability\ } q,\\  
\end{cases}
\end{equation}
where $p+q=1$.  If $k=j$, the winner is chosen randomly.  Initially,
all teams have zero wins and zero losses.

We use a kinetic framework to analyze the outcome of this random
process \cite{book}, taking advantage of the fact that the number of
games is a measure of time.  We randomly choose the two competing
teams and update the time by \hbox{$t\to t+\Delta t$}, with \hbox{$\Delta
t=1/(2N)$}, after each competition.  With this normalization, each team
participates in one competition per unit time.

Let $f_k(t)$ be the fraction of teams with $k$ wins at time $t$. This
probability distribution must be normalized, $\sum_k f_k =1$.  In the limit
$N\to \infty$, this distribution evolves according to
\begin{align}
\begin{split}
\label{fk-eq}
\frac{df_k}{dt}
&=p(f_{k-1}F_{k-1}-f_kF_k)\\
&+q(f_{k-1}G_{k-1}-f_kG_k)
+\frac{1}{2}(f_{k-1}^2-f_k^2)\,,
\end{split}
\end{align}
for $k\geq 0$.  Here we also introduced two cumulative distribution
functions: $F_k=\sum_{j=0}^{k-1} f_j$ is the fraction of teams with less than
$k$ wins and $G_k=\sum_{j=k+1}^{\infty} f_j$ is the fraction of teams with
more than $k$ wins. Of course, $F_k+G_{k-1}=1$.  The first two terms on the
right-hand-side of \eqref{fk-eq} account for games in which the stronger team
wins, and the next two terms correspond to matches where the weaker team
wins. The last two terms account for games between teams of equal strength
(the numerical prefactor is combinatorial).  Accounting for the boundary
condition $f_{-1}\equiv 0$ and summing the rate equations \eqref{fk-eq}, we
readily verify that the normalization $\sum_k f_k=1$ is preserved. The
initial conditions are $f_k(0)=\delta_{k,0}$.

In contrast to $f_k$, the cumulative distribution functions obey closed
evolution equations.  In particular, the quantity $F_k$ evolves according to
\cite{bvr}
\begin{eqnarray}
\label{Fk-eq}
\frac{dF_k}{dt}=q(F_{k-1}-F_k)+\left(\tfrac{1}{2}-q\right)\left(F_{k-1}^2-F_k^2\right),
\end{eqnarray}
which may be obtained by summing \eqref{fk-eq}. The boundary
conditions are $F_0=0$ and $F_\infty=1$, and the initial condition is
$F_k(0)=1$ for $k>0$.  We note that the average number of wins,
$\langle k\rangle =t/2$, where $\langle k\rangle =\sum_k kf_k$,
follows from the fact that each team participates in one competition
per unit time and that one win is awarded in each game.  As
\hbox{$\langle k\rangle =\sum_k k(F_{k+1}-F_k)$}, we can verify that
\hbox{$d\langle k\rangle/dt =1/2$} by summing the rate equations
\eqref{Fk-eq}.

We first discuss the asymptotic behavior when the number of games is
very large.  In the limit $t\to\infty$, we use the continuum approach 
and replace the difference equations \eqref{Fk-eq} with the partial
differential equation \cite{gbw,jmb}
\begin{equation}
\label{F-eq}
\frac{\partial F}{\partial t}+\big[q-(1-2q)F\big]\frac{\partial F}{\partial k}=0\,.
\end{equation}
According to our model, the weakest team wins at least a fraction $q$ of its
games, on average, and similarly, the strongest team wins no more than a
fraction $p$ of its games.  Hence, the number of wins is proportional to
time, $k \sim t$.  We thus seek the scaling solution
\begin{equation}
\label{Phi-def}
F_k(t)\simeq \Phi\left(\frac{k}{t}\right).
\end{equation}
Here and throughout this paper, the quantity $\Phi(x)$ is the scaled
cumulative distribution of win percentage; that is, the fraction of teams
that win less than a fraction $x$ of games played.  The boundary conditions
are $\Phi(0)=0$ and $\Phi(\infty)=1$.

We now substitute the scaling form \eqref{Phi-def} into \eqref{F-eq},
and find that the scaling function satisfies
\hbox{$\Phi'[(x-q)-(1-2q)\Phi]=0$} where prime denotes derivative with
respect to $x$. There are two solutions: \hbox{$\Phi={\rm constant}$}
and the linear function \hbox{$\Phi=(x-q)/(1-2q)$}. Therefore, the
distribution of win percentages is piecewise linear
\begin{equation}
\label{Phi-sol}
\Phi(x)=
\begin{cases}
0 & 0\leq x\leq q,\\
\frac{x-q}{p-q}& q\leq x\leq p, \\
1& p\leq x. 
\end{cases}
\end{equation}
As expected, there are no teams with win percentage less than the upset
probability $q$, and there are no teams with win percentage greater than the
complementary probability $p$. Furthermore, one can verify that
\hbox{$\langle x\rangle =1/2$}.  The linear behavior in \eqref{Phi-sol}
indicates that the actual distribution of win percentage becomes uniform,
$\Phi'=1/(p-q)$ for $q<x<p$, when the number of games is very large.

\begin{figure}[t]
\includegraphics*[width=0.45\textwidth]{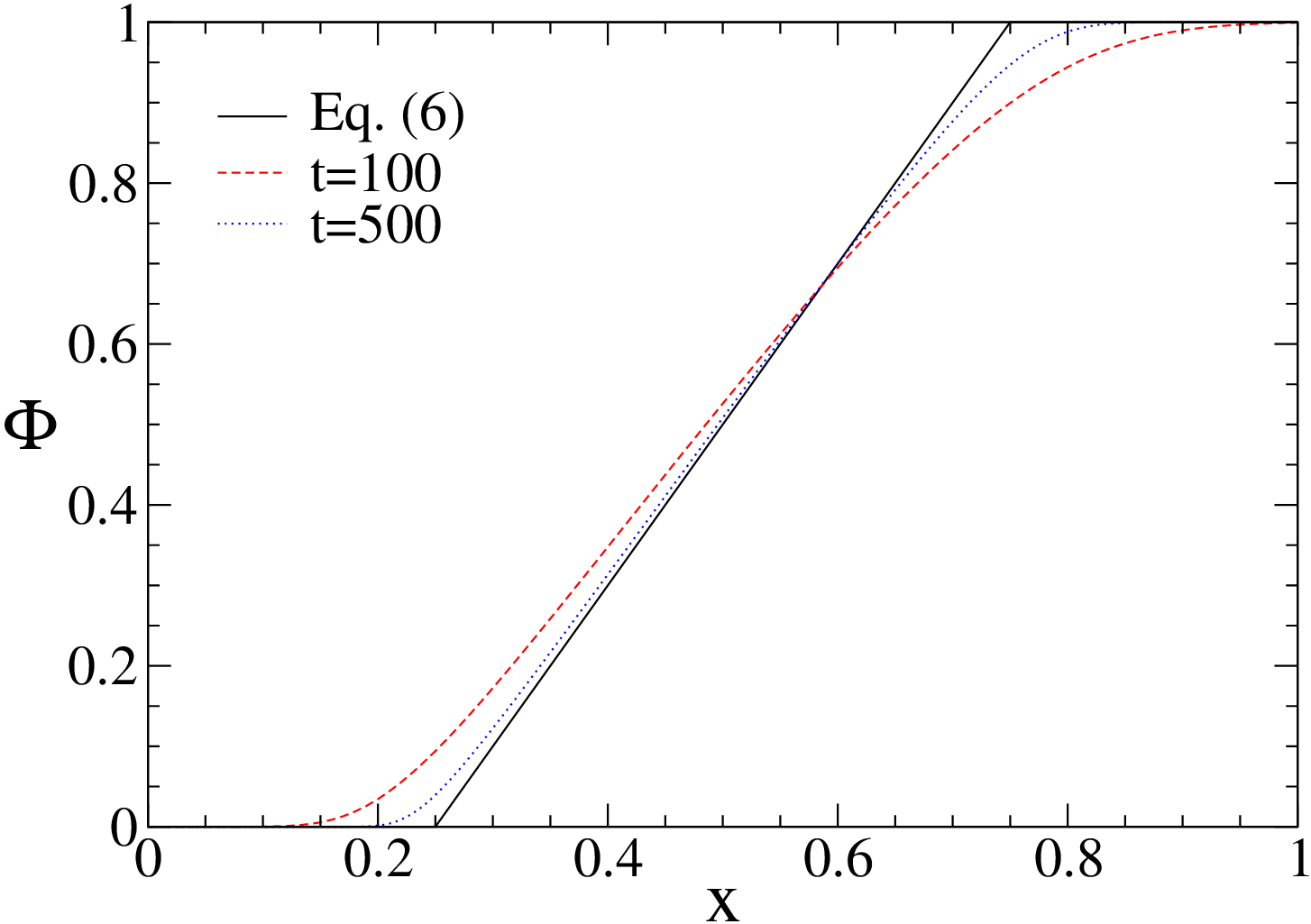}
\caption{The cumulative distribution $\Phi(x)$ versus win percentage
$x$ for $q=1/4$ at times $t=100$ and $t=500$. Also shown for reference
is the limiting behavior \eqref{Phi-sol}.}
\label{fig-phi}
\end{figure}

As shown in figure 1, direct numerical integration of the rate equation
\eqref{F-eq} confirms the scaling behavior \eqref{Phi-def}. Moreover, as the
number of games increases, the function $\Phi(x)$ approaches the
piecewise-linear function given by equation \eqref{Phi-sol}. However, there
is a diffusive boundary layer near $x=q$ and $x=p$, whose width decreases as
$t^{-1/2}$ in the long-time limit \cite{gbw}.

Generally, the win percentage is a convenient measure of team strength.  For
example, Major League Baseball (MLB) in the United States, where teams
play $\approx160$ games during the regular season, uses win percentage to
rank teams. The fraction of games won is preferred over the number of wins
because throughout the season there are small variations between the number
of games played by various teams in the league.

\begin{figure}[t]
\includegraphics*[width=0.45\textwidth]{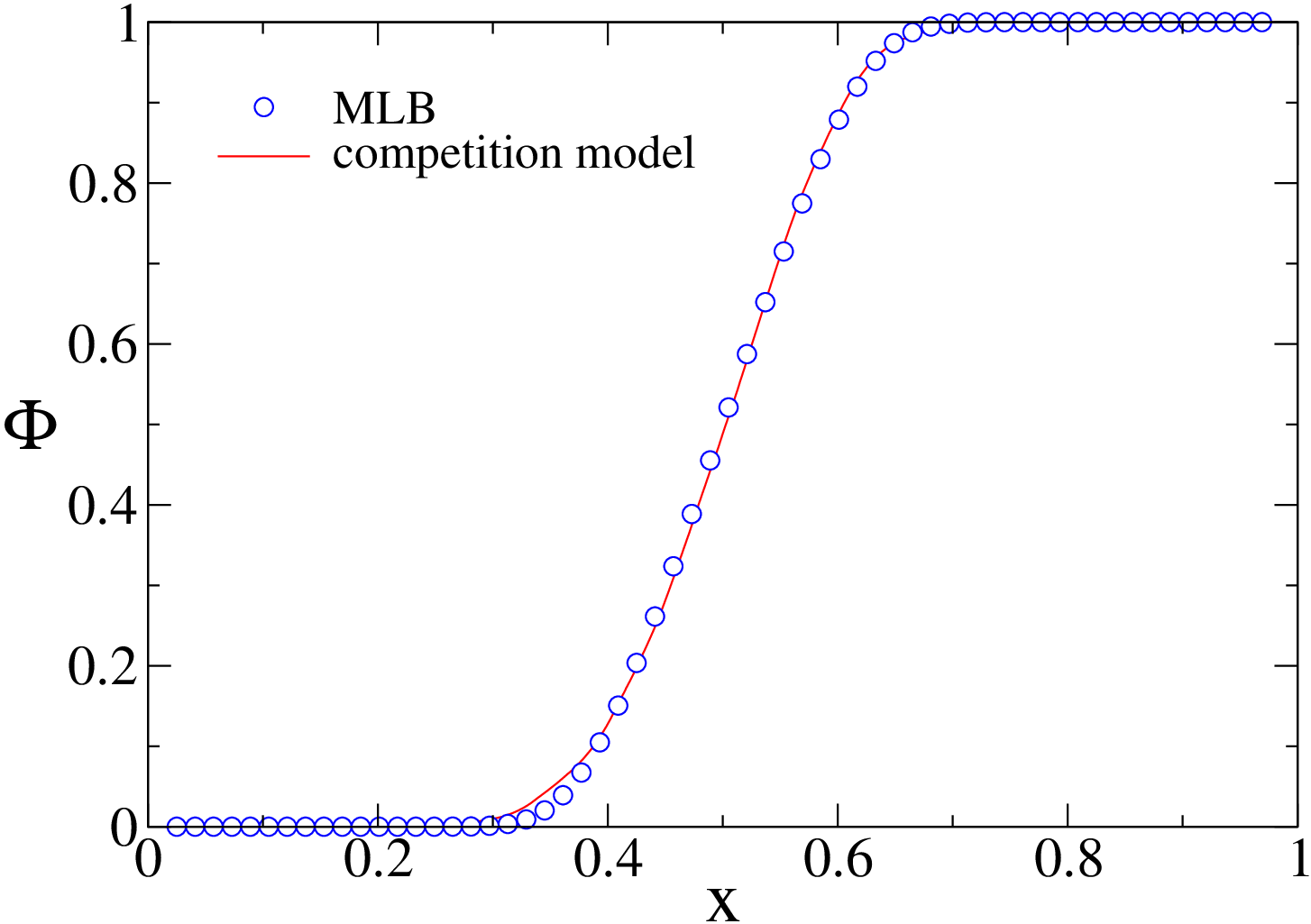}
\caption{The cumulative distribution $\Phi(x)$ versus win percentage
$x$ for: (i) Monte Carlo simulations of the competition process
\eqref{compete} with $q_{\rm model}=0.41$, and (ii) Season-end
standings for Major League Baseball (MLB) over the past century
(1901-2005).}
\label{fig-phi-mlb}
\end{figure}

The piecewise-linear scaling function in \eqref{Phi-sol} holds in the
asymptotic limits $N\to\infty$ and $t\to\infty$. To apply the
competition model \eqref{compete}, we must use a realistic number of
games and a realistic number of teams. To test whether the competition
model faithfully describes the win percentage of actual sports
leagues, we compared the results of Monte Carlo simulations with
historical data for a variety of sports leagues \cite{bvr1}. In this
paper, we give one representative example: Major League Baseball.

In our simulations, there are $N$ teams, each participating in exactly $t$
games throughout the season. In each match, two teams are selected at random,
and the outcome of the competition follows the stochastic rule
\eqref{compete}: with the upset probability $q$, the team with the lower win
percentage is victorious, but otherwise, the team with the higher win
percentage wins. At the start of the simulated season, all teams have an
identical record. We treated the upset frequency as a free parameter and
found that the value $q_{\rm model}=0.41$ best describes the historical data
for MLB ($N=26$ and $t=162$).  As shown in figure \ref{fig-phi-mlb}, the
competition model faithfully captures the empirical distribution of win
percentages at the end of the season. The latter distribution is calculated
from all season-end standings over the past century (1901-2005).

In addition, we directly measured the actual upset frequency $q_{\rm data}$
from the outcome of all $\approx 163,000$ games played over the past
century. To calculate the upset frequency, we chronologically ordered all
games and recreated the standings at any given day.  Then we counted the
number of games in which the winner was lower in the standings at the time of
the current game.  Game location and the margin of victory were ignored.  For
MLB, we find the value $q_{\rm data}=0.44$, only slightly higher than the
model estimate $q_{\rm model}=0.41$.

The standard deviation in win percentage, $\sigma$, defined by
$\sigma^2=\langle x^2\rangle-\langle x\rangle^2$, is commonly used to
quantify parity of a sports league \cite{fq,fm}.  For example, in
baseball, where the win percentage typically varies between $0.400$
and $0.600$, the historical standard deviation is $\sigma=0.084$.
From the cumulative distribution \eqref{Phi-sol}, it straightforwardly
follows that the standard deviation varies linearly with the upset
probability,
\begin{equation}
\label{sigma}
\sigma=\frac{1/2-q}{\sqrt{3}}.
\end{equation}
There is an obvious relationship between the predictability of
individual games and the competitive balance of a league: the more
random the outcome of an individual game, the higher the degree of parity
between teams in the league.

\begin{figure}[t]
\includegraphics*[width=0.45\textwidth]{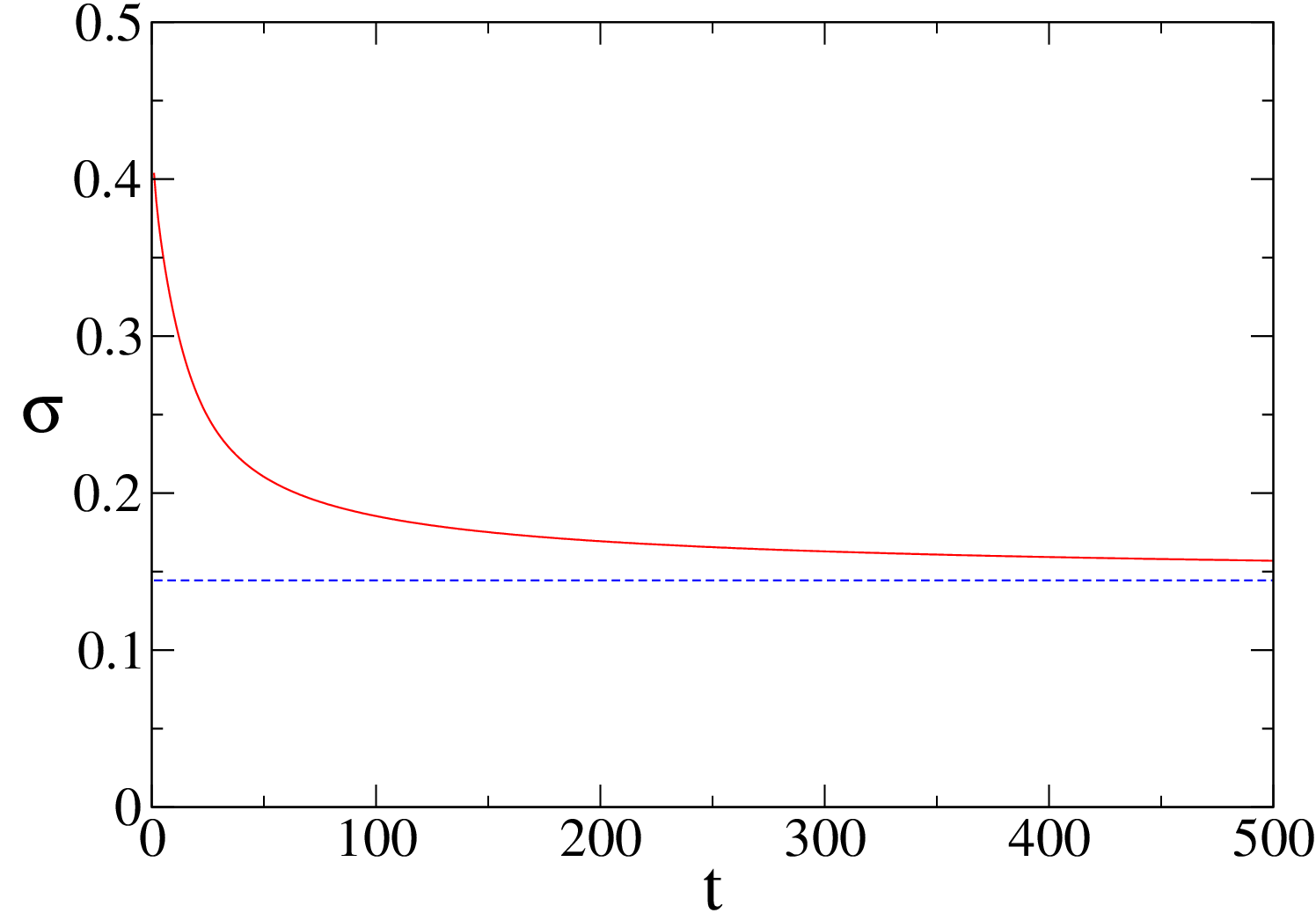}
\caption{The standard deviation $\sigma$ as a function of time
$t$. Shown are results of numerical integration of the rate equation
\eqref{fk-eq} with $q=1/4$. Also shown for reference is the limiting
value $\sigma_\infty=1/(4\sqrt{3})$.}
\label{fig-sigmat}
\end{figure}

The standard deviation is a convenient quantity because it requires
only year-end standings, which consist of only $N$ data points per
season. The upset frequency, on the other hand, requires the outcome of
each game, and therefore involves a much larger number of data points,
$Nt/2$ per season. Yet, as a measure for competitive balance, the
upset frequency has an advantage \cite{bvr1}. As seen in figure
\ref{fig-sigmat}, the quantity $\sigma$ consists of two contributions:
one due to the intrinsic nature of the game and one due to the finite
length of the season. For example, the large standard deviation
$\sigma=0.21$ in the National Football League (NFL) is in large part
due to the extremely short season, $t=16$. Therefore, the upset
frequency, which is decoupled from the length of the season, provides
a more accurate measure of competitive balance
\cite{jw,tl,hs1,hs2,hs3}.

\begin{figure}[t]
\includegraphics*[width=0.45\textwidth]{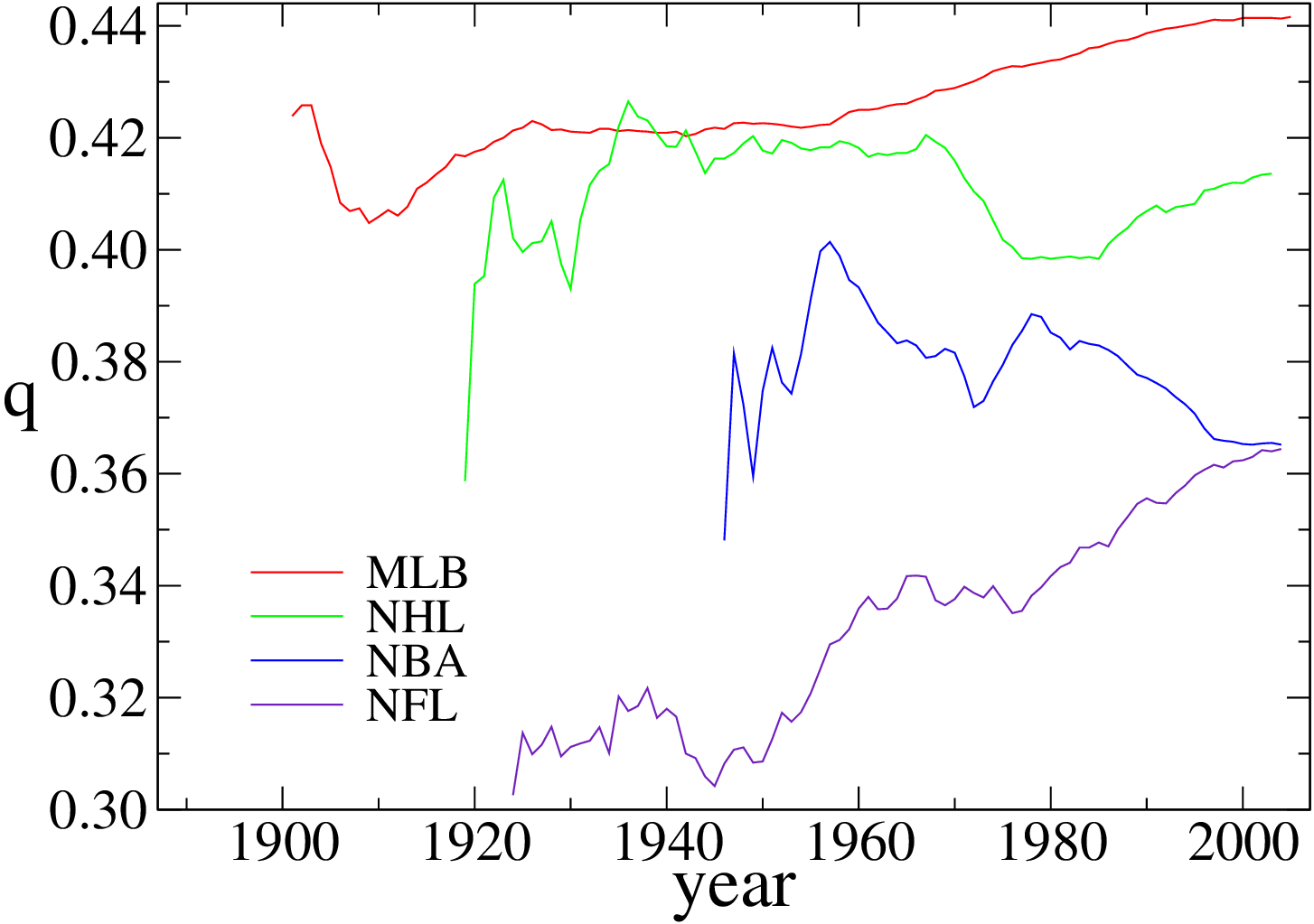}
\caption{ Evolution of the upset frequency $q$ with time. Shown is data
  \cite{data} for: (i) Major League Baseball (MLB), (ii) the National Hockey
  League (NFL) (iii) the National Basketball Association (NBA), and (iv) the
  National Football League (NFL).  The quantity $q$ is the cumulative upset
  frequency for all games played in the league up to the given year. In 
  football, a tie counts as one half of a win.}
\label{fig-sigma}
\end{figure}

The evolution of the upset frequency over time is truly fascinating
(figure \ref{fig-sigma}).  Although $q$ varies over a
narrow range, this quantity can differentiate the four sports leagues.
The historical data shows that MLB has consistently had the least
predictable games, while NBA and NFL games have been the most predictable.  
The trends for $q$ for these sports leagues are even more interesting.
Certain sports leagues (MLB and to a larger extent, NFL) managed to
increase competitiveness by changing competition formats, increasing
the number of teams, having unbalanced schedules where stronger teams
play more challenging opponents, or using a draft where the weakest
team can first pick the most promising upcoming talent.  

In spite of the fact that NHL and NBA implemented some of these same
measures to increase competitiveness, there are no clear long-term
trends in the evolution of the upset probability in these two leagues.
Another plausible interpretation of figure \ref{fig-sigma} is that the
sports leagues are striving to achieve an optimal upset frequency of
$q\approx 0.4$.  One may even speculate that the various sports
leagues compete against each other to attract public interest, and
that making the games less predictable, and hence, more interesting to
follow is a key objective in this evolutionary-like process
\cite{sjg,hs,lhn}.  In any event, the upset frequency is a natural and
transparent measure for the evolution of competitive balance in sports
leagues.

The random process \eqref{compete} involves only a single parameter, $q$. The
model does not take into account many aspects of real competitions including
the game score, the game location, the relative team strength, and the fact
that in many sports leagues the schedule is unbalanced, as teams in the same
geographical region may face each other more often.  Nevertheless, with
appropriate implementation, the competition model specified in equation
\eqref{compete} captures basic characteristics of real sports leagues. In
particular, the model can be used to estimate the distribution of team win
percentages as well as the upset frequency.

\section{Single Elimination Tournaments} 

Thus far, our approach did not include the notion of innate team
strength.  Randomness alone controlled which team reaches the top of the
standings and which teams reaches  at the bottom. Indeed, the
probability that a given team has the best record at the end of the
season equals $1/N$.  Furthermore, we have used the cumulative
win-loss record to define team strength. However, this definition can
not be used to describe tournaments where the number of games is
small.

We now focus on single-elimination tournaments, where the winner of a
game advances to the next round of play while the loser is eliminated
\cite{brv,fca}.  A single-elimination tournament is the most efficient
competition format: a tournament with $N=2^r$ teams requires only
$N-1$ games through $r$ rounds of play to crown a champion. In the
first round, there are $N$ teams and the $N/2$ winners advance to the
next round.  Similarly, the second round produces $N/4$ winners.  In
general, the number of competitors is cut by half at each round
\begin{equation}
\label{number}
N\to N/2 \to N/4\to \cdots \to 2 \to 1.
\end{equation}

In many tournaments, for example, the NCAA college basketball tournament in
the United States or in tennis championships, the competitors are ranked
according to some predetermined measure of their strength. Thus, we introduce
the notion of rank into our modeling framework. Let $x_i$ be the rank of the
$i$th team with
\begin{equation}
\label{rank}
x_1<x_2<x_3<\cdots<x_N.
\end{equation}
In our definition, a team with lower rank is stronger. Rank measures
innate strength, and hence, it does not change with time. Since
ranking is strict, we use the uniform ranking scheme $x_i=i/N$ without
loss of generality.

Again, we assume that there is a fixed probability $q$ that the
underdog wins the game, so that the outcome of each match is
stochastic. When a team with rank $x_1$ faces a team with rank $x_2$,
we have
\begin{equation}
\label{eliminate}
(x_1,x_2)\to
\begin{cases}
x_1\quad{\rm with\ probability\ } p,\\  
x_2\quad{\rm with\ probability\ } q,\\  
\end{cases}
\end{equation}
when $x_1<x_2$. The important difference with \eqref{compete} is that
the losing team is now eliminated.

Let $w_1(x)$ be the distribution of rank for all competitors. This
quantity is normalized, $\int_0^\infty dx\, w_1(x)=1$. In a two-team
tournament, the rank distribution of the winner, $w_2(x)$, is given by
\begin{equation}
\label{w2-eq}
w_2(x)=2p\,w_1(x)\left[1-W_1(x)\right]+2q\,w_1(x)W_1(x),
\end{equation}
where $W_1(x)=\int_0^x dy\, w_1(y)$ is the cumulative distribution of
rank. The structure of this equation resembles that of \eqref{fk-eq}, with
the first term corresponding to games where the favorite advances, and the
second term to games where the underdog advances.  Mathematically, there is a
basic difference with Eq.~\eqref{fk-eq} in that equation \eqref{w2-eq} does
not contain loss terms.  Again, ties are not allowed to occur.  By
integrating \eqref{w2-eq}, we obtain the closed equation
\hbox{$W_2(x)=2pW_1(x)+(1-2p)\big[W_1(x)\big]^2$}.

\begin{figure}[t]
\includegraphics*[width=0.45\textwidth]{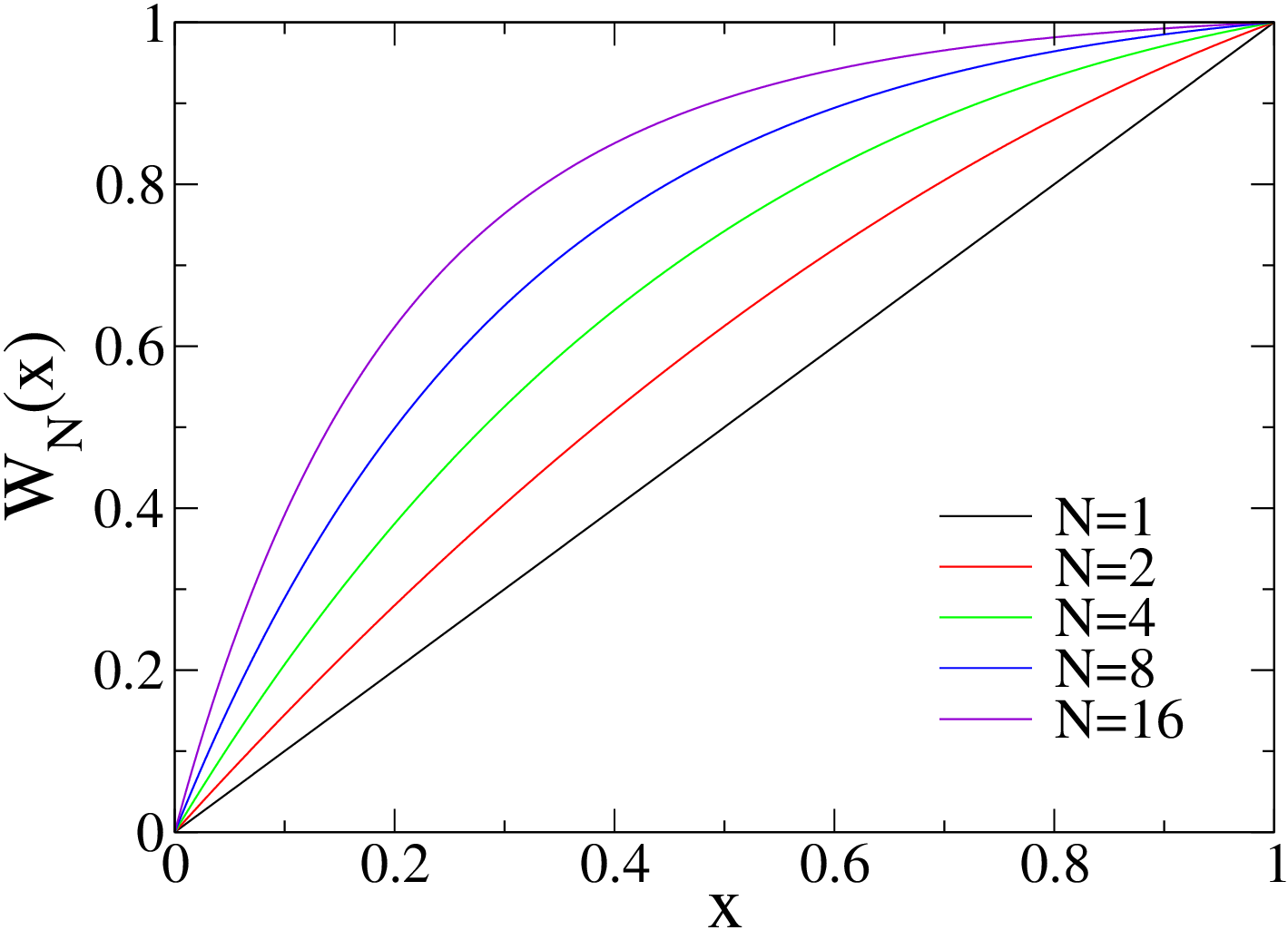}
\caption{The cumulative distribution of rank. The quantity $W_N(x)$
is calculated by iterating equation \eqref{wN-eq} with $q=1/4$.}
\label{fig-wn}
\end{figure}

In general, the cumulative distribution obeys the nonlinear recursion
equation
\begin{equation}
\label{wN-eq}
W_{2N}(x)=2pW_N(x)+(1-2p)\big[W_N(x)\big]^2.
\end{equation}
Here, $W_N(x)=\int_0^x dy\, w_N(y)$, and $w_N(x)$ is the rank distribution
for the winner of an $N$-team tournament.  The boundary conditions are
$W_N(0)=0$ and $W_N(\infty)=1$.  The prefactor $2$ arises because there are
two ways to choose the winner.  The quadratic nature of equation
\eqref{wN-eq} reflects that two teams compete in each match (competitions
with three teams are described by cubic equations \cite{bkk,mt,rt}). Starting
with $W_1(x)=x$ that corresponds to uniform ranking, $w_1(x)=1$, we can
follow how the distribution of rank evolves by iterating the recursion
equation \eqref{wN-eq}. As shown in figure \ref{fig-wn}, the rank of the
winner decreases as the size of the tournament increases.  Hence, larger
tournaments produce stronger winners.

By substituting $W_1(x)=x$ into equation \eqref{wN-eq}, we find
$W_2(x)=(2p)x$ and in general, $W_N(x)=(2p)^r\,x$. This behavior
suggests the scaling form
\begin{equation}
\label{WN-scaling}
W_N(x)\simeq \Psi(x/x_*),
\end{equation}
where the scaling factor $x_*$ is the typical rank of the winner. This
quantity decays algebraically with the size of the tournament, 
\begin{equation}
\label{x-scaling}
x_*=N^{-\beta},\qquad \beta=\frac{\ln (2p)}{\ln 2}.
\end{equation}
When games are perfectly random (upset probability \hbox{$q=1/2$}),
the typical rank of the winner becomes independent of the number of
teams, $\beta(q\!=\!1/2)=0$. When the games are highly predictable,
the top teams tend to win the tournament, $\beta(0)=1$. Again, the
scaling behavior \eqref{x-scaling} shows that larger tournaments tend
to produce stronger champions.

By substituting \eqref{WN-scaling} into \eqref{wN-eq}, we see that the
scaling function $\Psi(z)$ obeys the nonlocal and nonlinear equation
\begin{equation}
\label{Psi-eq}
\Psi(2pz)=2p\Psi(z)+(1-2p)\Psi^2(z).
\end{equation}
The boundary conditions are $\Psi(0)=0$ and $\Psi(\infty)=1$. From
equation \eqref{Psi-eq}, we deduce the asymptotic behaviors
\begin{equation}
\Psi(z)\simeq \begin{cases}
z & z\to 0, \\
1-C\,z^\gamma & z\to\infty,
\end{cases}
\end{equation}
with the scaling exponent $\gamma=\frac{\ln (2q)}{\ln (2p)}$. The
large-$z$ behavior is obtained by substituting $\Psi(z)=1-U(z)$
into \eqref{Psi-eq} and noting that since $U\to 0$ when $z\to\infty$,
the correction obeys the linear equation $U(2pz)=2qU(z)$.

The large-$z$ behavior of the scaling function $\Psi(z)$ gives the
likelihood that a very weak team manages to win the entire
tournament. The scaling behavior \eqref{WN-scaling} is equivalent to
$w_N(x)\simeq (1/x_*)\psi(x/x_*)$ with $\psi(z)=\Psi'(z)$. In the limit
$z\to 0$, the distribution approaches a constant $\psi(z)\to
1$. However, the tail of the rank distribution is algebraic
\begin{equation}
\label{powerlaw}
\psi(z)\sim z^{-\alpha},\qquad \alpha=1-\frac{\ln (2q)}{\ln (2p)},
\end{equation}
when $z\to\infty$.  The exponent $\alpha>1$ increases monotonically
with $p$, and it diverges in the limit $p\to 1$ \cite{deterministic}.

Moreover, the probability that the weakest team wins the tournament,
$P_N=q^N$, decays algebraically with the total number of teams,
$P_N=N^{\ln q/\ln 2}$.  In the following section, we discuss sports
leagues and find that: (i) the rank distribution of the winner has an
{\it exponential} tail, and (ii) the probability that the weakest team
is crowned league champion is exponentially small.

The scaling behavior \eqref{WN-scaling} indicates universal statistics
when the size of the tournament is sufficiently large. Once rank is
normalized by typical rank, the resulting distribution does not depend
on tournament size.  Further, the scaling law \eqref{x-scaling} and
the power-law tail \eqref{powerlaw} reflect that tournaments can
produce major upsets.  With a relatively small number of upset wins, a
``Cinderella'' team can emerge, and for this reason, tournaments can
be very exciting.  Furthermore, tournaments are maximally efficient as
they require a minimal number of games to decide a champion.

\begin{figure}[t]
\includegraphics*[width=0.45\textwidth]{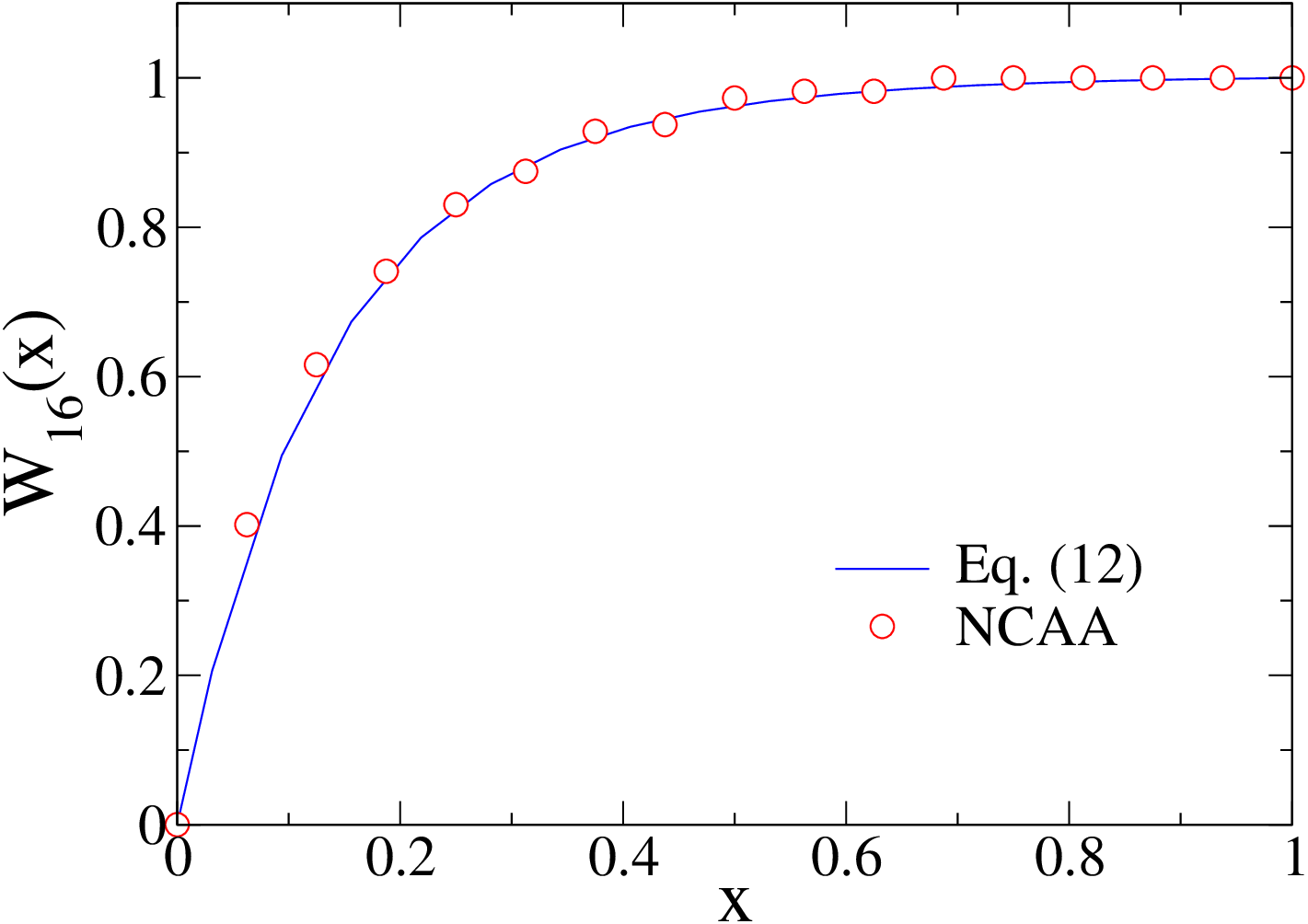}
\caption{The cumulative distribution of rank for the NCAA college
basketball tournament. Shown is the cumulative distribution
$W_{16}(x)$ versus the rank $x$ for (i) NCAA tournament data
(1979-2006), (ii) Iteration of the equation \eqref{wN-eq}.}
\label{fig-ncaa}
\end{figure}

Figure \ref{fig-ncaa} shows that our theoretical model nicely
describes empirical data \cite{data} for the NCAA college basketball
tournament in the United States \cite{brv}. In the current format, 64
teams participate in four sub-tournaments, each with $N=16$ teams. The
four winners of each sub-tournament advance to the final four, which
ultimately decides the champion.  Prior to the tournament, a committee
of experts ranks the teams from $1$ to $16$. We note that the game
schedule is not random, and is designed such that the top teams
advance if there are no upsets.

Consistent with our theoretical results, the NCAA tournament has been
producing major upsets: the $11$th seed team has advanced to the final
four twice over the past 30 years. Moreover, only once did all of the
four top-seeded teams advance simultaneously (2008).  Our model
estimates the probability of this event at $1/190$, a figure that is
of the same order of magnitude as the observed frequency $1/132$.

We also mention that in producing the theoretical curve in figure
\ref{fig-ncaa}, we used the upset frequency \hbox{$q_{\rm
    model}=0.18$}, whereas the actual game results yield $q_{\rm
  data}=0.28$. This larger discrepancy (compared with the MLB analysis
above) is due to a number of factors including the much smaller
dataset ($\approx 7000$ games) and the non-random game
schedule. Indeed, our Monte-Carlo simulations which incorporate a
realistic schedule give better estimates for the upset frequency
\cite{brv}.

\section{Leagues}

We now discuss the common competition format in which each team hosts every
other team exactly once during the season.  This format, first used in
English soccer, has been adopted in many sports.  In a league of size $N$,
each team plays $2(N-1)$ games and the total number of games equals
$N(N-1)$. Given this large number of games, does the strongest team always
wins the championship?

To answer this question, we assume that each team has an innate
strength and rank the teams according to strength.  Without loss of
generality, we use the uniform rank distribution $w(x)=1$ and its cumulative
counterpart $W(x)=x$ where $0\leq x\leq 1$. Moreover, we implicitly
take the large-$N$ limit.  Consider a team with rank $x$.  The
probability $v(x)$ that this team wins a game against a
randomly-chosen opponent decreases linearly with rank,
\begin{equation}
\label{vx}
v(x)=p-(2p-1)x,
\end{equation}
as follows from $v(x)=p[1-W_1(x)]+qW_1(x)$ [see also equation
  \eqref{w2-eq}]. Consistent with our competition rules
\eqref{compete} and \eqref{eliminate}, the probability $v(x)$
satisfies $q\leq v\leq p$.  

Since team strength does not change with time, the average number of
wins $V(x,t)$ for  a team with rank $x$ grows linearly with the
number of games $t$,
\begin{equation}
\label{Vt}
V(x,t)=v(x)\,t. 
\end{equation}
Accordingly, the number of wins of a given team performs a biased random
walk: after each game the number of wins increases by one with probability
$v$, and remains unchanged with the complementary probability $1-v$. Also,
the uncertainty in the number of wins, $\Delta V$, grows diffusively with
$t$,
\begin{equation}
\label{DeltaVt}
\Delta V(x,t)\simeq \sqrt{Dt},
\end{equation}
with diffusion coefficient $D=v(1-v)$ \cite{book}.

Let us assume that each team plays $t$ games. If the number of games
is sufficiently large, the best team has the most wins.  However, at
intermediate times, it is possible that a weaker team has the most
wins.  For a team with strength $x_*$ to still be in contention at
time $t$, the difference between its expected number of wins and that
of the top team should be comparable with the diffusive uncertainty
\begin{equation}
\label{Delta}
V(0,t)-V(x_*,t)\sim \Delta V(0,t).
\end{equation}
We now substitute equations \eqref{vx}-\eqref{DeltaVt} into this heuristic
estimate and obtain the typical rank of the leader as a function of
time, 
\begin{equation}
\label{xt}
x_*\sim \frac{1}{\sqrt{t}}.
\end{equation}
In obtaining this estimate, we tacitly ignored numeric prefactors,
including in particular, the dependence on $q$.

This crude estimate \eqref{xt} shows that the best team does not always win
the league championship.  Since $t\sim N$, we have
\begin{equation}
\label{xN}
x_*\sim \frac{1}{\sqrt{N}}.
\end{equation}
Since rank is a normalized quantity, the top $\sqrt{N}$ of the teams have a
realistic chance of emerging with the best record at the end of the season.
Thus randomness plays a crucial role in determining the champion: since the
result of an individual game is subject to randomness, the outcome of a long
series of games reflects this randomness.

\begin{figure}[t]
\includegraphics*[width=0.45\textwidth]{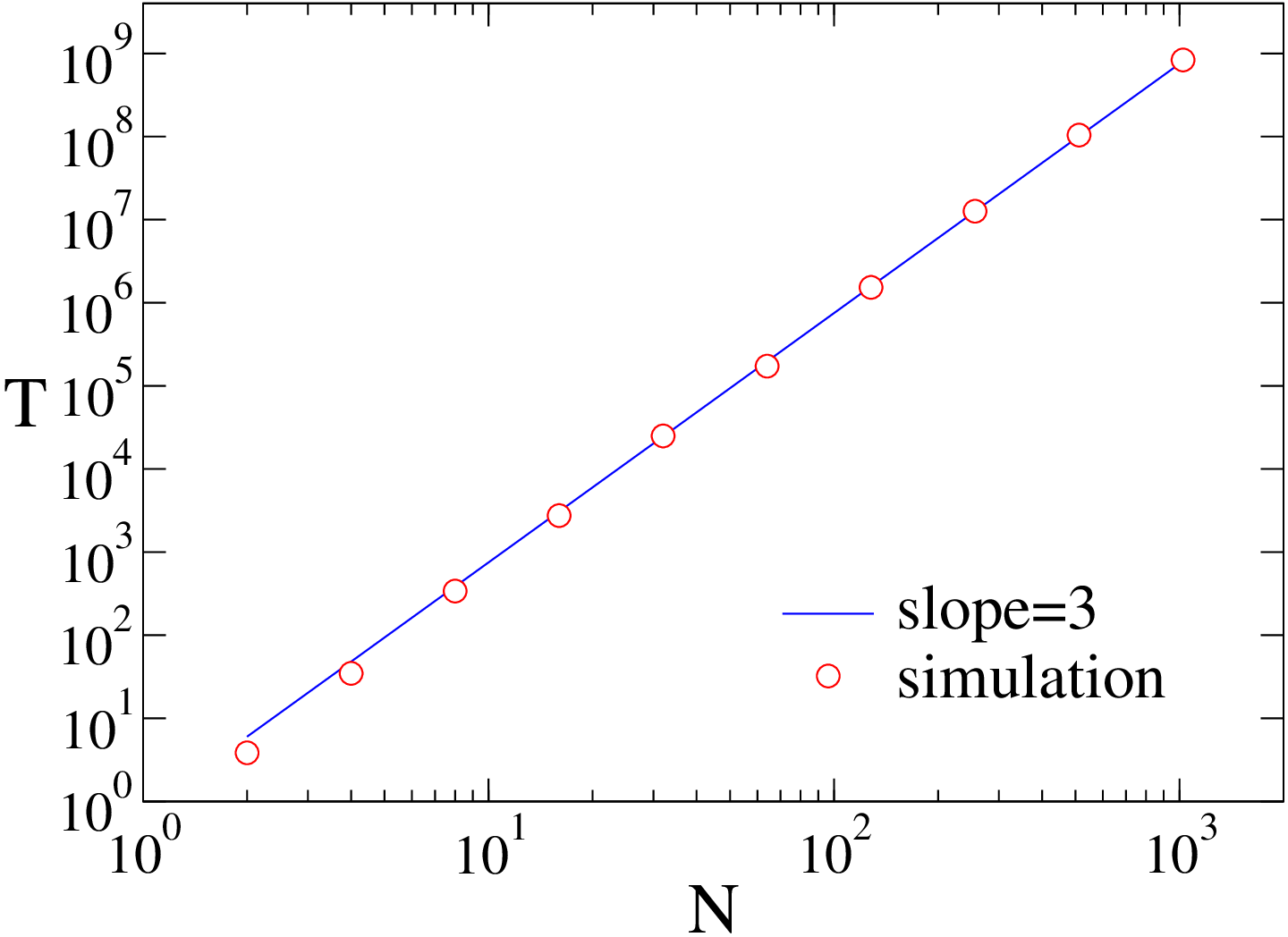}
\caption{The total number of games $T$ needed for the best team to
emerge as champion in a league of size $N$. The simulation results
represent an average over $10^3$ simulated sports leagues. Also shown
for  reference is the theoretical prediction.}
\label{fig-TN}
\end{figure}

We can also obtain the total number of games $T$ needed for the best
team to always emerge as the champion, 
\begin{equation}
\label{TN}
T\sim N^3.
\end{equation}
This scaling behavior follows by replacing $x_*$ in \eqref{xt} with $1/N$
which corresponds to the best team. For the best team to win, each team must
play every other team ${\cal O}( N)$ times!  Alternatively the number of
games played by each team scales quadratically with the size of the
league. Clearly, such a schedule is prohibitively long, and we conclude that
the traditional schedule of playing each opponent with equal frequency is
neither efficient nor does it guarantee the best champion.

We confirmed the scaling law \eqref{TN}  numerically. In
our Monte Carlo simulations, the teams are ranked from $1$ to $N$ at the
start of the season. We implemented the traditional league format
where every team plays every other team and kept track of the leader
defined as the team with the best record. We then measured the
last-passage time \cite{sr}, that is, the time in which the best team
takes the lead for good. We define the average of this fluctuating
quantity as $T$ \cite{kr,bk}.  As shown in figure \ref{fig-TN}, the
total number of games required is cubic.

Again, we expect that the probability distribution $w(x,t)$ that a
team with rank $x$ has the best record after $t$ games is
characterized by the scale $x_*$ given in \eqref{xt}
\begin{equation}
\label{wxt-scaling}
w(x,t)\simeq (1/x_*)\varphi(x/x_*).
\end{equation}
Numerical results confirm this scaling behavior \cite{bh}.  Since the
number of wins performs a biased random walk, we expect that the
distribution of the number of wins becomes normal in the long-time
limit.  Moreover, the scaling function in \eqref{wxt-scaling} has a 
Gaussian tail \cite{bh}
\begin{equation}
\label{varphi}
\varphi(z)\sim \exp\left(-{\rm const.}\times z^2\right),
\end{equation}
as $z\to \infty$. 

Using this scaling behavior, we can readily estimate the probability
that worst team becomes champion (in the standard league format). For
the worst team, $x\sim 1$, and the corresponding scaling variable in
equation \eqref{wxt-scaling} is $z\sim \sqrt{N}$. Hence, the Gaussian
tail \eqref{varphi} shows that the probability $P_N$ that the weakest
team wins the league is exponentially small,
\begin{equation}
\label{exponential}
P_N\sim \exp\left(-{\rm const.}\times N\right).
\end{equation}
In sharp contrast with tournaments, where this probability is
algebraic, leagues do not produce upset champions. Leagues may not
guarantee the absolute top team as champion, but nevertheless,
they do produce worthy champions.

\begin{figure}[t]
\includegraphics*[width=0.45\textwidth]{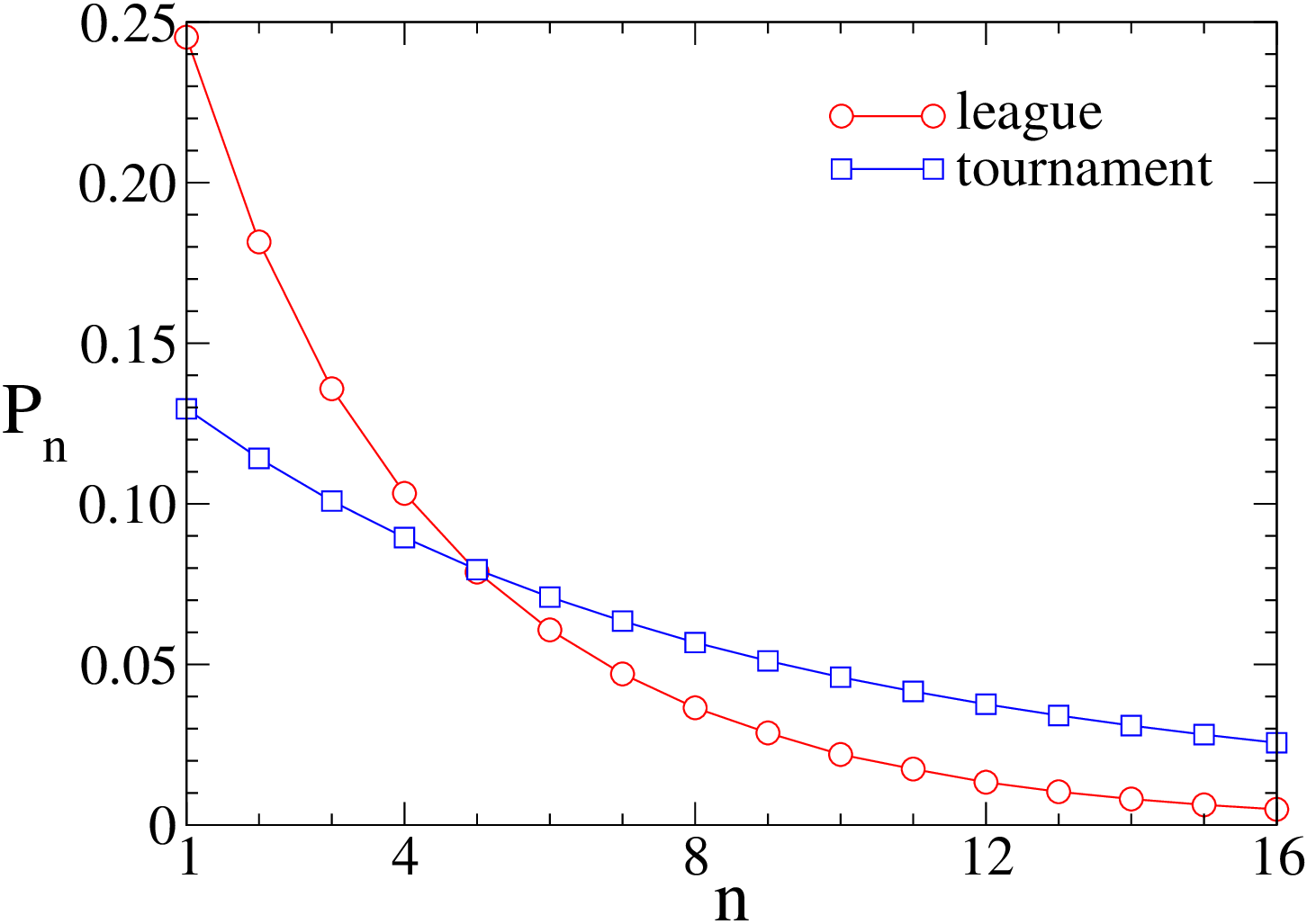}
\caption{Leagues versus tournaments. Shown is $P_n$, the probability that the
  $n^{\rm th}$-ranked team has the best record at the end of the season in
  the format of playing all opponents with equal frequency, and the
  probability that the $n^{\rm th}$-ranked team wins an $N$-team
  single-elimination tournament. The upset probability is $q=0.4$ and
  $N=16$.}
\label{fig-lt}
\end{figure}

To compare leagues and tournaments, we calculated the probability
$P_n$ that the $n$th ranked team is champion for a realistic number of
games $N=16$ and a realistic upset probability $q=0.4$ (figure
\ref{fig-lt}). For leagues, we calculated this probability from Monte
Carlo simulations, and for tournaments, we used equation
\eqref{wN-eq}. Indeed, the top four teams fare better in a league
format while the rest of the teams are better off in a
tournament. This behavior is fully consistent with the above estimate
that the top $\sqrt{N}$ teams have a realistic chance to win the
league.

What is the probability $P_{\rm top}$ that the top team ends the
season with the best record in a realistic sports league? To answer
this question, we investigated the four major sports leagues in the
US: MLB, NHL, NFL, and NBA. We simulated a league with the actual
number of teams $N$ and the actual number of games $t$, using the
empirical upset frequencies (see figure \ref{fig-sigmat}). All of
these sports leagues have comparable number of teams, $N\approx 25$.
Surprisingly, we find almost identical probabilities for three of the
sports leagues: (i) MLB with the longest season and most random games
($t=162$, $q=0.44$) has $P_{\rm top}=0.31$, (ii) NFL with the shortest
season but most deterministic games ($t=16$, $q=0.37$) has $P_{\rm
top}=0.30$, and (iii) NHL with intermediate season and intermediate
randomness ($t=80$, $q=0.41$) has $P_{\rm top}=0.32$.  Standing out as
an anomaly is the value $P_{\rm top}=0.45$ for the NBA which has a
moderate-length season but less random games ($t=80$ and $q=0.37$).

This interesting result reinforces our previous comments about sports leagues
competing against each other for interest and our hypothesis that there are
optimal randomness parameters.  Having a powerhouse win every year does not
serve the league well, but having the strongest team finish with the best
record once every three years may be optimal.

\section{Gradual Elimination Algorithm}

Our analysis demonstrates that single-elimination tournaments have
optimal efficiency but may produce weak champions, whereas leagues
which result in strong winners are highly inefficient. Can we devise a
competition ``algorithm'' that guarantees a strong champion within a
minimal number of games?

As an efficient algorithm, we propose a hybrid schedule consisting of
a preliminary round and a championship round \cite{bh}. The
preliminary round is designed to weed out a majority of teams using a
minimal number of games, while the championship round includes ample
games to guarantee the best team wins.

In the preliminary round, every team competes in $t$ games. Whereas
the league schedule has complete graph structure with every team
playing every other team, the preliminary round schedule has regular
random graph structure with each team playing against the same number
of randomly-chosen opponents.  Out of the $N$ teams, the $M$ teams
with the largest number of wins in the preliminary-round advance to
the championship round. The number of games $t$ is chosen such that
the strongest team always qualifies.  By the same heuristic argument
\eqref{Delta} leading to \eqref{xt}, the top team ranks no lower than
$1/\sqrt{t}$ after $t$ games. We thus require
\begin{equation}
\frac{M}{N}\sim \frac{1}{\sqrt{t}},
\end{equation}
and consequently, each team plays $\sim(N/M)^2$ preliminary games. The
championship round uses a league format with each of the $M$
qualifying teams playing $M$ games against every other
team. Therefore, the total number of games, $T$, has two components
\begin{equation}
\label{T1}
T\sim \frac{N^3}{M^2}+M^3.
\end{equation}
In writing this estimate, we ignore numeric prefactors, as well as the
dependence on the upset frequency $q$. The quantity $T$ is minimal
when the two terms in \eqref{T1} are comparable \cite{dg}. Hence, the
size of the championship round $M_1$ and the total number of games
$T_1$ scale algebraically with $N$,
\begin{equation}
\label{M1T1}
M_1\sim N^{3/5},\quad {\rm and}\quad T_1\sim N^{9/5}.
\end{equation}
Consequently, each team plays ${\cal O}\big(N^{4/5}\big)$ games in the
preliminary round.  Interestingly, the existence of a preliminary round
significantly reduces the number of games from $N^3$ to $N^{9/5}$.  Without
sacrificing the quality of the champion, the hybrid schedule yields a huge
improvement in efficiency!

We can further improve the efficiency by using multiple elimination
rounds. In this generalization, there are $k-1$ consecutive rounds of
preliminary play culminating in the championship round. The underlying
graphical structure of the preliminary rounds is always a regular
random graph, while the championship round remains a complete
graph. Each preliminary round is designed to advance the top teams,
and the number of games is sufficiently large so that the top team
advances with very high probability. When there are $k$ rounds, we
anticipate the scaling laws
\begin{equation}
\label{MT}
M_k\sim N^{\nu_k},\quad {\rm and}\quad T_k\sim N^{\mu_k},
\end{equation}
where $M_k$ is the number of teams advancing out of the first round
and $T_k$ is the total number of games.  Of course, when there are no
preliminary rounds, $\nu_0=1$ and $\mu_0=3$.  Following equation
\eqref{MT}, the number of teams gradually declines in each round,
\begin{equation}
N\to N^{\nu_k}\to N^{\nu_k\nu_{k-1}}\to \cdots \to  N^{\nu_k\nu_{k-1}\cdots\nu_1}\to 1.
\end{equation}

\begin{table}[t]
\begin{tabular}{|c|c|c|c|c|c|c|c|}
\hline
\,$k$\,&\,0\,&\,1\,&\,2\,&\,3\,&\,4\,&$\,\infty$\,\\
\hline
$\nu_k$&0&$\frac{3}{5}$&$\frac{15}{19}$&$\frac{57}{65}$&$\frac{195}{211}$&1\\
$\mu_k$&3&$\frac{9}{5}$&$\frac{27}{19}$&$\frac{81}{65}$&$\frac{243}{211}$&1\\
\hline
\end{tabular}
\caption{The exponents $\nu_k$ and $\mu_k$ in equation \eqref{MT} for
$k\leq 4$.}
\end{table}

According to the first term in \eqref{T1}, the number of games in the
first round scales as $N^3/M_k^2\sim N^{3-2\nu_k}$, and
therefore, the total number of games obeys the recursion
\begin{equation}
\label{Tk-eq}
T_k\sim N^{3-2\nu_k}+T_{k-1}(N^{\nu_k}).
\end{equation}
Indeed, if we replace $M_1$ with $N^{\nu_1}$ in equation \eqref{T1}
we can recognize the recursion \eqref{Tk-eq}.  The second term scales
as $N^{\nu_k\mu_{k-1}}$ and becomes comparable to the second when
$3-2\nu_k=\nu_k\mu_{k-1}$. Hence, the scaling exponents satisfy the
recursion relations
\begin{equation}
\label{nu-eq}
\nu_k=\frac{3}{2+\mu_{k-1}},\quad {\rm and}\quad \mu_k=\mu_{k-1}\nu_k.
\end{equation}
Using $\nu_0=1$ and $\mu_0=3$, we recover $\nu_1=3/5$ and $\mu_1=9/5$
in agreement with \eqref{M1T1}.  The general solution of \eqref{nu-eq}
is \cite{bh}
\begin{equation}
\label{numu}
\nu_k=\frac{1-\left(2/3\right)^k}{\,\,\,\,\,1-\left(2/3\right)^{k+1}}, \qquad 
\mu_k=\frac{1}{1-(2/3)^{k+1}}.
\end{equation}
Hence, the efficiency is optimal, and the number of games becomes linear
in the limit $k\to\infty$. For a modest number of teams, a small
number of preliminary rounds, say 1-3 rounds, may suffice. Indeed,
with as few as four elimination rounds, the number of games becomes
essentially linear, $\mu_4\cong 1.15$.

Interestingly, the result $\mu_\infty=1$ indicates that championship
rounds or ``playoffs'' have the optimal size $M_*$ given by 
\begin{equation}
\label{opt}
M_*\sim N^{1/3}.
\end{equation}
Gradual elimination is often used in the arts and sciences to decide
winners of design competitions, grant awards, and prizes.  Indeed, the
selection process for prestigious prizes typically begins with a quick
glance at all nominees to eliminate obviously weak candidates, but
concludes with rigorous deliberations to select the winner.  Multiple
elimination rounds may be used when the pool of candidates is very
large.

To verify numerically the scaling laws \eqref{M1T1}, we simulated a
single preliminary round followed by a championship round. We chose
the size of the preliminary round strictly according to \eqref{MT} and
used a championship round where all $M_1$ teams play against all $M_1$
teams exactly $M_1$ times. We confirmed that as the number of teams
increases from $N=10^1$ to $10^2$ to $10^3$ etc., the probability that
the best team emerges as champion is not only high but also,
independent of $N$.  We also confirmed that the concept of preliminary
rounds is useful for small $N$. For $N=10$ teams, the number of games
can be reduced by a factor $>10$ by using a single preliminary round.

\section{Discussion}

We introduced an elementary competition model in which a
weaker team can upset a stronger team with fixed probability. The model
includes a single control parameter, the upset frequency, a quantity
that can be measured directly from historical game results. This idealized
competition model can be conveniently applied to a variety of
competition formats including tournaments and leagues. The random
competition process is amenable to theoretical analysis and is
straightforward to implement in numerical simulations.  Qualitatively,
this model explains how tournaments, which require a small number of
games, can produce major upsets, and how leagues which require a large
number of games always produce quality champions. Additionally, the
random competition process enables us to quantify these intuitive features:
the rank distribution of the champion is algebraic in the former schedule 
but Gaussian in the latter.

Using our theoretical framework, we also suggested an efficient algorithm
where the teams are gradually eliminated following a series of preliminary
rounds. In each preliminary round, the number of games is sufficient to
guarantee that the best team qualifies to the next round. The final
championship round is held in a league format in which every team plays many
games against every other team to guarantee that the strongest team emerges
as champion. Using gradual elimination, it is possible to choose the champion
using a number of games that is proportional to the total number of
teams. Interestingly, the optimal size of the championship round scales as
the one third power of the total number of teams.

The upset frequency plays a major role in our model. Our empirical studies
show that the frequency of upsets, which shows interesting evolutionary
trends, is effective in differentiating sports leagues. Moreover, this
quantity has the advantage that it is not coupled to the length of the
season, which varies widely from one sport to another. Nevertheless, our
approach makes a very significant assumption: that the upset frequency is
fixed and does not depend on the relative strength of the
competitors. Certainly, our approach can be generalized to account for
strength-dependent upset frequencies \cite{sr1}. We note that our
single-parameter model fares better when the games tend to be close to
random, and that model estimates for the upset frequency have larger
discrepancies with the empirical data when the games become more
predictable. Clearly, a more sophisticated set of competition rules are
required when the competitors are very close in strength, as is the case for
example, in chess \cite{meg}.

We thank Micha Ben-Naim for help with data collection. We acknowledge
support from DOE (DE-AC52-06NA25396) and NSF (DMR0227670, DMR0535503,
\& DMR-0906504).


\begin{thebibliography}{99}

\bibitem{csv}
         C.~Castellano, S.~Fortunato, and V.~Loreto, 
         Rev.~Mod.~Phys {\bf 81}, 591 (2009).

\bibitem{btd} 
         E.~Bonabeau, G~Theraulaz, and J.-L.~Deneubourg, 
         Physica A {\bf 217}, 373 (1995).

\bibitem{br} 
         E.~Ben-Naim and S.~Redner
         J. Stat.\ Mech.\ L11002 (2005).

\bibitem{msk}
         K.~Malarz, D.~Stauffer, K.~Kulakowski,
         Eur.\ Phys.\ Jour.\ B {\bf 50}, 195 (2006).

\bibitem{ra}
         R.~L.~Axtell, Science {\bf 293}, 5536(2001).

\bibitem{sjg} 
         S.~J.~Gould,
         {\sl Full house: The spread of excellence from Plato to Darwin} 
         (Harmony Books, New York, 1996).

\bibitem{pn} 
         J.~Park and M.~E.~J.~Newman, 
         J. Stat.\ Mech.\ P10014 (2005).

\bibitem{pjs}
         A.~M.~Petersen, W.~S.~Jung, H.~E.~Stanley, 
         Europhys.\ Lett.\ {\bf 83}, 50010 (2008).

\bibitem{hmr}
         A.~Heuer, C.~Mueller, and O.~Rubner,
         Europhys.\ Lett.\ {\bf 89}, 38007 (2010).

\bibitem{fr}
         F.~Radicchi,
         PLOS ONE {\bf 6}, e17249 (2011).

\bibitem{abc} 
       J.~Albert, J.~Bennett,  and J.~J.~Cochran, eds.\ 
       {\sl Anthology of Statistics in Sports} 
       (SIAM, Philadelphia, 2005).

\bibitem{gts} 
       D.~Gembris, J.~G.~Taylor, and D.~Suter, 
       Nature {\bf 417}, 506 (2002).

\bibitem{bvr} 
       E.~Ben-Naim, F.~Vazquez, and S.~Redner,
       Eur.\ Phys.\ Jour.\ B {\bf 26}, 531 (2006).

\bibitem{bvr1} 
       E.~Ben-Naim, F.~Vazquez, and S.~Redner,
       J. Quant.\ Anal.\ Sports {\bf 2}, No. 4, Article 1 (2006);
       J. Korean Phys.\ Soc.\ {\bf 50}, 124 (2007).

\bibitem{brv}
         E.~Ben-Naim, S.~Redner, and F.~Vazquez, 
         Europhys.\ Lett.\ {\bf 77}, 30005 (2007).

\bibitem{bh} 
         E.~Ben-Naim and N.~W.~Hengartner, 
         Phys.\ Rev.\ E {\bf 76}, 026106 (2007).

\bibitem{book}  
         P.~L.~Krapivsky, S.~Redner and E.~Ben-Naim,
         {\sl  A Kinetic View of Statistical Physics}
         (Cambridge University Press, Cambridge, UK, 2010).

\bibitem{gbw}
         G.~B.~Whitham,
         {\sl Linear and Nonlinear Waves},
         (Wiley, New York, 1974).

\bibitem{jmb} 
         J.~M.~Burgers,
         {\sl The nonlinear diffusion equation}
         (Reidel, Dordrecht, 1974).

\bibitem{fq} 
         R.~Fort and J.~Quirk,  
         J. Econ.\ Liter.\ {\bf 33}, 1265 (1995).

\bibitem{fm} 
         R.~Fort and J.~Maxcy,  
         J. Sports Econ.\ {\bf 4}, 154 (2003).

\bibitem{jw} 
         J.~Wesson, 
         {\sl The Science of Soccer} 
	 (IOP, Bristol  and Philadelphia, 2002).

\bibitem{tl} 
         T.~Lundh,  
         J. Quant.\ Anal.\ Sports {\bf 2}: No. 3, Article 1 (2006).

\bibitem{hs1} 
        H.~S.~Stern, 
        The American Statistician {\bf 45}, 179 (1991).

\bibitem{hs2} 
         H.~S.~Stern, H.S., 
         Chance {\bf 10}, 19 (1997).

\bibitem{hs3} 
          H.~S.~Stern and B.~R.~Mock, 
          Chance {\bf 11}, 26 (1998).

\bibitem{data} 
         Data source: http://www.shrpsports.com/.

\bibitem{hs} 
         J.~Hofbauer and K.~Sigmund,  
         {\sl Evolutionary Games and Population Dynamics} 
         (Cambridge University Press, Cambridge, 1998).

\bibitem{lhn} 
         E.~Lieberman, Ch.~Hauert, and M.~A.~Nowak, 
         Nature {\bf 433}, 312 (2005).

\bibitem{fca}
         T.~M.~A.~Fink, J.~B.~Coe, and S.~E.~Ahnert, 
         Europhys.\ Lett.\ {\bf 83}, 60010 (2008).

\bibitem{bkk} 
         E.~Ben-Naim, B.~Kahng, and J.~S.~Kim
         J. Stat.\ Mech.\ P07001 (2006)

\bibitem{mt}
         M.~Mungan and T.~Rador, 
         J.~Phys.~A {\bf 41},  055002 (2008).

\bibitem{rt}
         T.~Rador and R.~Derici, 
         Eur.\ Phys.\ Jour.\ B {\bf 83}, 289 (2011). 

\bibitem{deterministic} 
          For deterministic competitions, $q=0$, the scaling function is
exponential $\psi(z)=e^{-z}$.

\bibitem{sr} 
         S.~Redner,
         {\sl A Guide to First-Passage Processes} 
         (Cambridge University Press, Cambridge 2001)

\bibitem{kr}
         P.~L.~Krapivsky and S.~Redner,
         Phys.\ Rev.\ Lett.\ {\bf 89}, 258703 (2002).

\bibitem{bk}
         E. Ben-Naim and P.~L.~Krapivsky
	 Europhys.\ Lett.\ {\bf 65}, 151 (2004) 

\bibitem{dg} 
         P.~G. de Gennes, 
        {\sl Scaling Concepts in Polymer Physics} 
        (Cornell University Press, Ithaca, 1979).  

\bibitem{sr1}
        C.~Sire and S.~Redner, 
        Eur.\ Phys.\ Jour.\ B {\bf 67}, 473 (2009).

\bibitem{meg}
        M.~E.~Glickman,
        American Chess Jour.\ {\bf 3}, 59 (1995).

\end{thebibliography}
\end{document}